\documentclass[preprint,3p]{elsarticle}
\usepackage{graphicx}
\usepackage{upgreek}

\newcommand{\degrees}[0]{$^\circ$}
\newcommand{\centigrade}[0]{^\circ \mathrm{C}}
\journal{Journal of Magnetic Resonance}
\begin{document}
\begin{frontmatter}
\title{Magnetic properties of materials for MR engineering, micro-MR and beyond}
\author[IMTEK]{Matthias C Wapler}
\author[UKF]{Jochen Leupold}
\author[UKF]{Iulius Dragonu}
\author[UKF]{Dominik von Elverfeld}
\author[UKF]{Maxim Zaitsev}
\author[IMTEK]{Ulrike Wallrabe}\ead{wallrabe@imtek.uni-freiburg.de}
\address[IMTEK]{Laboratory for Microactuators, Department of Microsystems Engineering - IMTEK, University of Freiburg, Germany}
\address[UKF]{Medical Physics, Department of Radiology, University Medical Center Freiburg, Germany}
\begin{abstract}
We present the results of a systematic measurement of the magnetic susceptibility of small material samples in a 9.4 T MRI scanner. We measured many of the most widely used materials in MR engineering and MR micro technology, including various polymers, optical and substrate glasses, resins, glues, photoresists, PCB substrates and some fluids. Based on our data, we identify particularly suitable materials with susceptibilities close to water. For polyurethane resins and elastomers, we also show the MR spectra, as they may be a good substitute for silicone elastomers and good casting resins.\\[5mm]
NOTICE: this is the author’s version of a work that was accepted for publication in Journal of Magnetic Resonance. Changes resulting from the publishing process, such as peer review, editing, corrections, structural formatting, and other quality control mechanisms may not be reflected in this document. Changes may have been made to this work since it was submitted for publication. A definitive version was subsequently published in Journal of Magnetic Resonance 242C (2014), pp. 233-242  DOI: 10.1016/j.jmr.2014.02.005
\end{abstract}
\end{frontmatter}
\section{Introduction}
Having a homogeneous magnetic field is essential for the majority of magnetic resonance experiments. Hence, the distortion of the magnetic field is an important design parameter in the construction of any apparatus that is in the vicinity of the sample during a measurement, including the sample itself. Problems arising from field distortions include line broadening or distortion in NMR spectroscopy and image distortions and signal loss in MR imaging \cite{schenck1996role,doty,fuks,ludeke1985susceptibility,callaghan1990susceptibility,drain1962broadening,pruessmann}. Thus, one of the most important material properties in MR engineering is the magnetic susceptibility. 
%
%
%

The largest collection of susceptibility data in the literature is the CRC handbook \cite{haynes2012crc} which, however, contains only elements and simple chemical compounds but not most polymers and glasses used in engineering. Beyond this, there are various measurements of smaller sets of sample materials obtained with different methods \cite{keyser1989magnetic,doty,fuks,pruessmann,hoffman,marcon2011magnetic}. Hence, we believe that a database of magnetic susceptibility values for commonly used materials in MR engineering and micro-MR, obtained in a consistent manner, would be a useful tool in the field. Additional values can be found in particular for metals in \cite{doty} for polymers in \cite{doty,keyser1989magnetic} and for fluids in \cite{fuks,pruessmann}, albeit usually at a lower accuracy.

In the present paper, we present the measurement results of a wide selection of polymers, PCB substrates, glues, photoresists, glasses and fluids. To test the variation within one type of material, we measured samples from different batches, manufacturers and colours for some exemplary materials. We also verified the agreement with the literature and the reproducibility of our results.
Particular focus is placed on polyurethane. This is available as an elastomer that can be used as an MR-compatible replacement for PDMS (silicone), and also as a rigid material that can be used as a casting resin. For this type of material, we studied also the effects of mixtures and variations of the processing conditions, and we show the MR spectrum for some representative samples. 
\section{Method, data analysis and error treatment}\label{methods}
A well-known method to measure the susceptibility of materials that do not produce an MR signal is to place the measurement samples in a reference material (for example water) with a known susceptibility and measure the field distortion in the reference material outside the sample. Typically, the sample is a long cylinder, for which the distortion is a dipole \cite{park1988measurement,beuf1996magnetic,marcon2011magnetic,liu2009calculation,carlsson2006accurate}. \cite{doty} uses a slightly special method, as there, the authors place two rings of the sample material in a Helmholtz configuration, such that the field distortion in the centre between the rings is homogeneous. They then measure the magnetic field at the centre using an NMR sample.  In \cite{fuks}, in contrast, the authors do not produce a homogeneous magnetic field but obtain the susceptibility by comparison of the distortion of NMR lines caused by different materials near the NMR sample.
For materials that do produce a signal, for example aqueous solutions and other fluids, another method is to measure the field distortion inside the sample \cite{beuf1996magnetic,wang1999magnetic,weis1994measurement,weisskoff1992mri,hoffman,pruessmann}. Careful construction and magnetic design of the probe \cite{pruessmann} and knowledge of the shape factor \cite{hoffman} are very important in this context.

We used the former method, but we designed it for relatively straightforward preparation of the samples and for independence from the positioning of the sample and for some degree of independence from accurate alignment. While the method of \cite{doty} is tolerant to overall displacements of the Helmholtz pair, it is still extremely sensitive to relative displacements and to the preparation of the sample.
Our strategy was to use samples with circular and quadratic cross sections which do simplify the sample preparation. We then performed a finite element simulation of the field distortion of finite-length samples that are oriented orthogonal to the magnetic field. This was compared to a measured 3D field map.
One simulation was sufficient for all samples with the same aspect ratio but with different overall size and different susceptibility because electromagnetism is linear and scale-invariant.
It was also sufficient to assume that the magnetisation of the samples is induced by a constant magnetic field because the relative field distortions are of the order of magnitude of the susceptibility and thus much smaller than our relative measurement accuracy.

Liquid samples were filled into small glass tubes and then treated similar to solid samples. In this case, we first measured the glass tube filled with the same reference material as the phantom. Tubes were modelled by the difference of two rods with different diameter. Then, the susceptibility value of the reference liquid inside was obtained and taken as a reference for the measured liquid in the other tubes. Typically this value was zero within errors, representing a good model and fitting.

\begin{figure}\begin{center}
\includegraphics[width=0.48\textwidth]{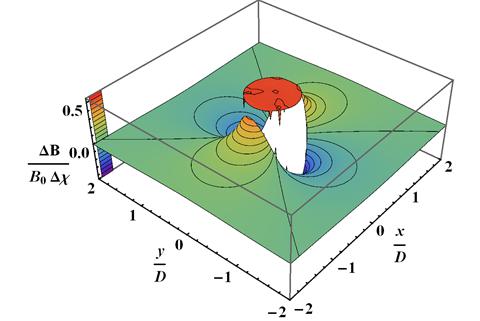}\hspace{0.02\textwidth}
\includegraphics[width=0.48\textwidth]{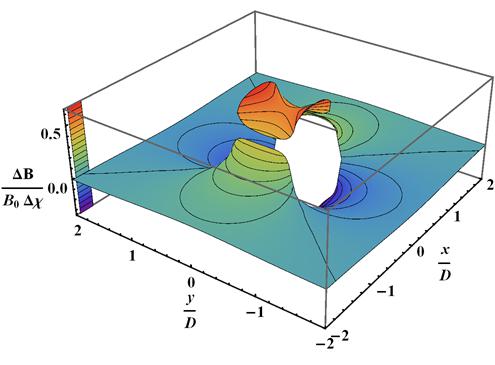}\hspace{0.02\textwidth}
\end{center}\caption{The distortion of the magnetic field in the central plane of a sample, normalised to the magnetic field and the susceptibility. The region at the edge of the sample was removed both for clarity of the figure and to illustrate the voxels that were ignored in the fit. The dimensions are normalised to the sample diameter. Left: Circular cross section. Right: Quadratic cross section.}\label{simfig} 
\end{figure}

We chose the aspect ratio of the samples around 10 as such samples are most convenient to handle. In fig. \ref{simfig}, we show the field distortion at the centre of such a bar, both for a cylindrical and a quadratic cross section. Furthermore, taking the length of the bars as a sample separation was sufficient to reduce the distortions due to neighbouring samples to background level. 

The data of the field distortion was obtained with a Bruker BioSpec 94/21 imaging system with $B_0 \, = \, 9.4\, \mathrm{T}$ and a BGA 12S gradient system with maximum gradient amplitude of 676 mT/m and 4750 mT/m/s maximum slew rate. For small samples, we used a quadrature birdcage Tx/Rx coil with 38 mm inner diameter and for large samples a linear polarised birdcage Tx/Rx coil with 72 mm inner diameter.
The field maps were obtained at a resolution of typically less than $1/10$ of the sample diameter.

To analyse the data, we first divided the field map into slices in coronal orientation, orthogonal to the sample. Depending on the position along the sample, the appropriate slice of the simulation was taken and an interpolation of the simulation result was produced. This interpolation was then fitted to the data by linear scaling, adjusting the central position of the sample and adding a linear or quadratic background field to account for imperfect shimming. The diameter of the sample was measured prior to the measurement and the dimensions of the simulation result were scaled to match the sample. 

To enhance the reliability of the measurement, all pixels inside the sample were removed from the data analysis, also in the cases where there appeared some signal.
The pixels at the edge of the sample may be affected by partial volume effects which may increase the noise and may yield an incorrect value for the field distortion in the corresponding points. Geometrically, they also cannot be clearly associated with either the sample material or the surrounding water. As the data analysis is, however, very sensitive to those pixels, we also removed them from the analysis. 

The uncertainty associated with the susceptibility of each slice was taken from the accuracy of the cross section of the sample. This represents the mechanical measurement, the asymmetry of the cross section, the roughness of the surface and the unevenness over the length of the sample. To model the influence on the susceptibility value, we simply scaled the dimensions of the simulation result in either direction with plus or minus the thickness uncertainty and obtained the RMS deviation of the fitted susceptibility values.
\begin{figure}\begin{center}
\includegraphics[width=0.31\textwidth]{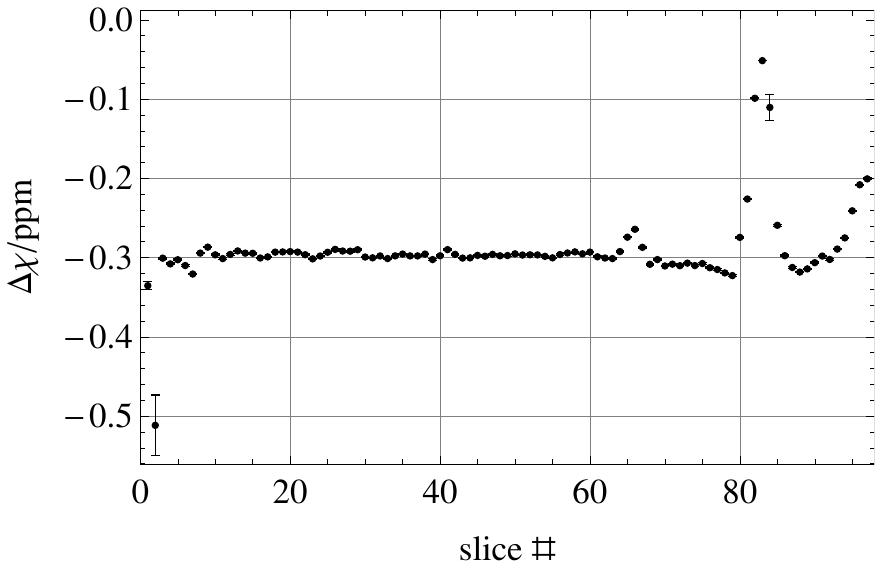}\hspace{0.02\textwidth}
\includegraphics[width=0.31\textwidth]{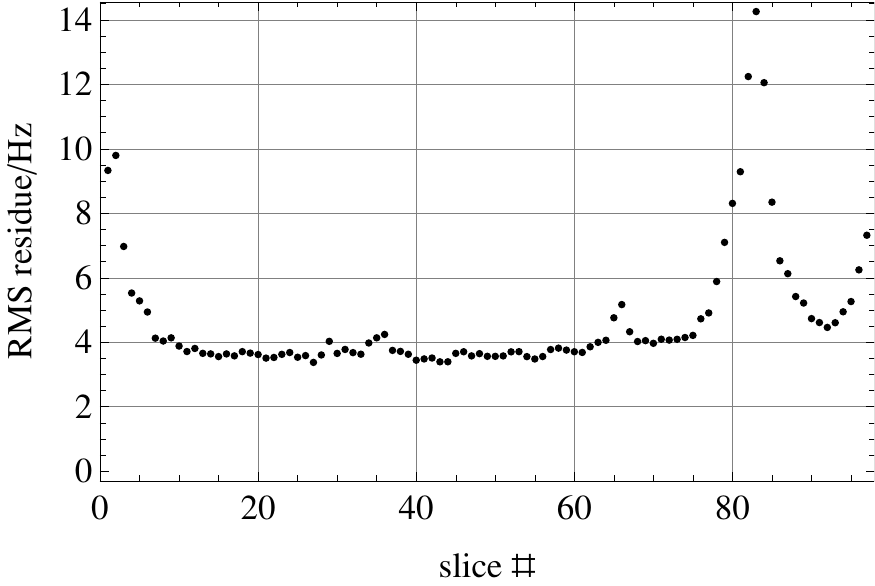}\hspace{0.02\textwidth}
\includegraphics[width=0.31\textwidth]{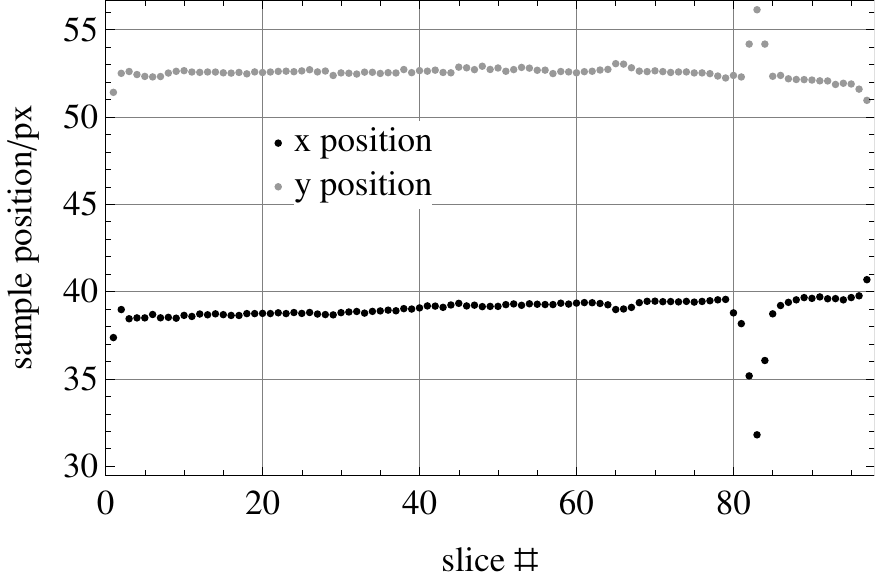}\hspace{0.02\textwidth}\\
\includegraphics[width=0.31\textwidth]{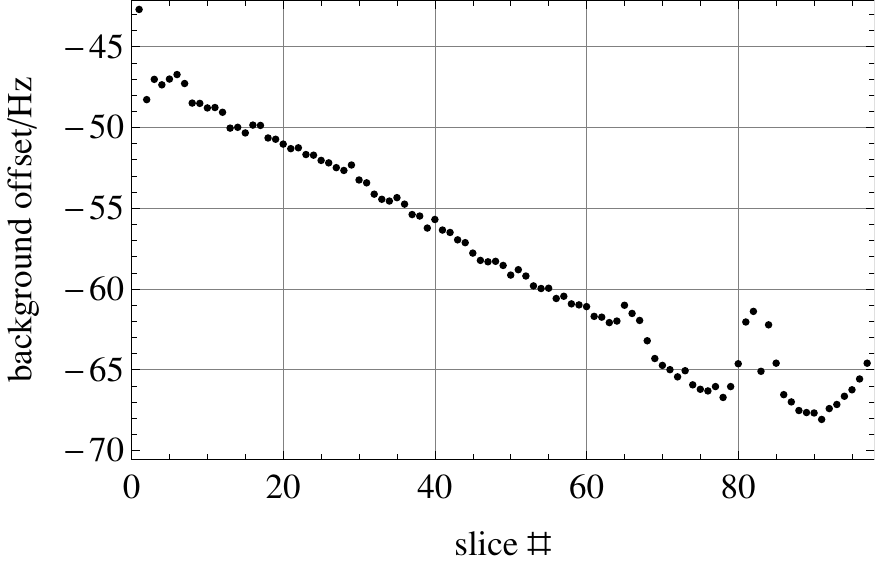}\hspace{0.02\textwidth}
\includegraphics[width=0.31\textwidth]{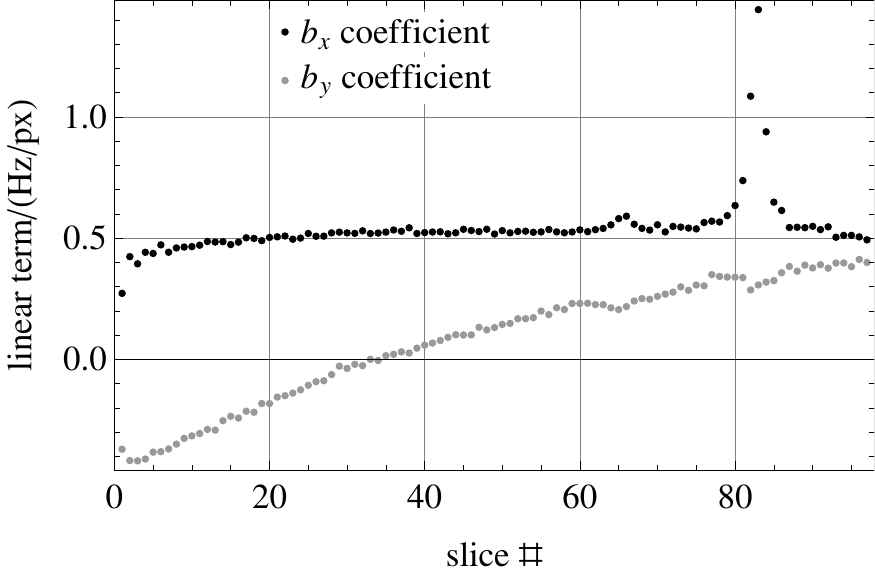}\hspace{0.02\textwidth}
\includegraphics[width=0.31\textwidth]{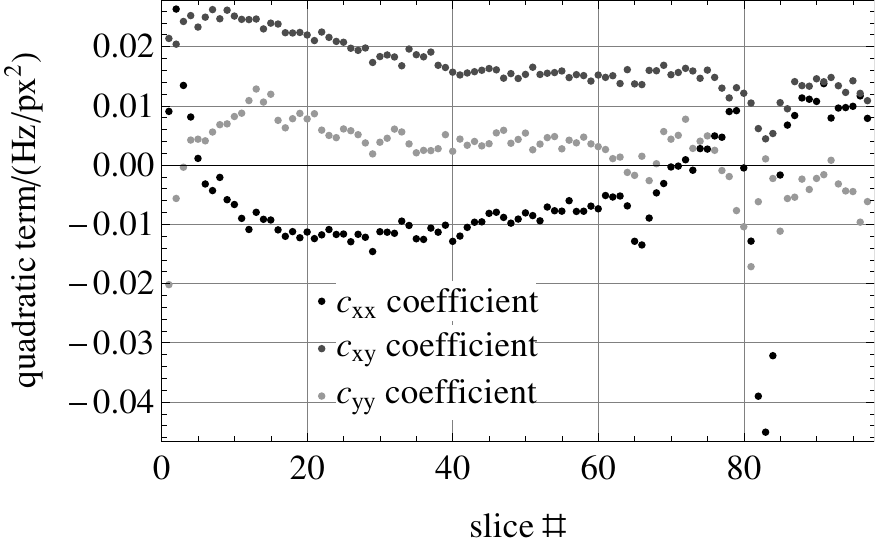}\hspace{0.02\textwidth}\\
\includegraphics[width=0.44\textwidth]{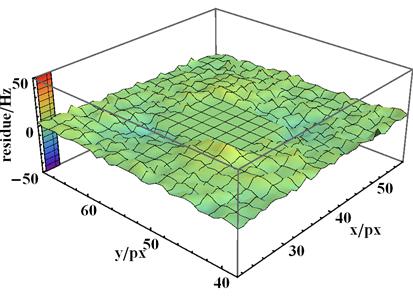}\hspace{0.05\textwidth}
\includegraphics[width=0.44\textwidth]{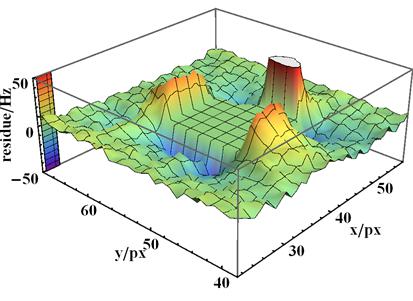}\hspace{0.05\textwidth}\\
\end{center}\caption{Top: Fitted susceptibility of PEEK (difference to water) at each slice along the sample, RMS residue of the fitted field map and fitted central position of the sample. Centre: Coefficients of the background fit of the field map $\Delta \nu_{\mathrm{background}} = a +b_x x + b_y y + c_{xx} x^2+c_{xy} xy + c_{yy}y^2$. Bottom: Residual map of slice 22 (left) and 83 (right). All positions are expressed in pixels and the magnetic field in frequency (Hz).}\label{fitfig} 
\end{figure}

The susceptibility values corresponding to the different slices were then filtered in several ways. Firstly, the RMS value of the residuals of the field map was used as a numerical cutoff. The cutoff was typically set at 2 to 5 times the value of the best slice, depending on the level of the systematic background deviation. In addition, the residual maps and the series of values obtained for the position and background parameters were inspected to manually filter out mis-fittings that may come from air bubbles or faults in the samples. This is illustrated in fig. \ref{fitfig}, where we show the sequence of susceptibility fits, RMS residuals, position and background fits for a slightly worse than usual sample (PEEK), one typical acceptable residual map and a non-acceptable one containing artifacts. There, we can actually see the artifacts of what appears to be an air bubble around slices 80 to 85, and some other artifacts or a smaller air bubble around slices 65 to 67. Furthermore, we see errors near the ends of the bar, and in the ``good'' residual map, we see a small rotation of the (quadratic) sample.

The remaining values are then averaged in a weighted mean using the uncertainty values of the susceptibility as weights. Strictly speaking, this suppresses values that are very sensitive to the cross section of the sample. Each sample has then two uncertainty values, one from the statistical distribution of the values of the different slices, and one from the mean (as they are systematic) 
of the uncertainties coming from the sample cross section, which are combined as if they were independent error sources. 

To further track and isolate the influence of the background distortion and possible aliasing from the background model, two different cross sections of the volume of interest were obtained from the data -- one $\sim 5$ times the sample diameter and one $\sim 7$ times the sample diameter. Both were then fitted with linear and quadratic backgrounds. As we had no quantitative measure for the quality of the background fit and for possible mis-fitting of data with the quadratic background parameters, we discarded up to two of these 4 data sets after visual inspection of the residual maps. The small field of view with the linear background model and the large field with the quadratic background model were kept in most cases. The resulting set of two to four susceptibility figures gave us another uncertainty value that expresses the sensitivity to the background distortion and background fitting.

What has not been taken into account are the variations in the composition of the samples. We assume that the manufacturing uncertainties of the material are much smaller than our measurement accuracy, and for some materials we tested this using materials from different sources or different batches. Furthermore, we did not continuously monitor the temperature. 
Monitoring during the measurement would lead to problems of how to associate measured voxels to temperatures and then correcting for the temperature changes, so we just take the maximum temperature variation as a basis for an additional error source. As the molar susceptibility is typically  independent of small temperature changes the volumetric susceptibility changes approximately with the thermal volume expansion \cite{schenck1996role,haynes2012crc,philo1980temperature}. The dominant thermal volume expansion coefficient for our relative measurement was typically the one of water with a value 
of $2\times 10^{-4} \mathrm{K}^{-1}$.
To estimate the resulting uncertainties, we measured the temperatures in the preparation room in which the samples were stored, in the room of the MRI device and the temperatures of a phantom before and after a 40 minutes scan, that was longer than the typical scan times by a factor 2 to 4. 
This gave us a temperature range of $20\pm 2 \, ^{\circ} \mathrm{C}$, resulting in an uncertainty in the susceptibility of $\pm 0.004 \mathrm{ppm}$.  As we did not measure any materials that are known to have a negative thermal expansion coefficient, even an expansion twice as large as the one of water does not change this error band.

Finally, the uncertainty from the temperature variation was added in quadrature to the other uncertainties.
\subsection{Sample preparation}
The samples were prepared in different ways. Solid samples that were already in cylindrical form with a diameter in the appropriate range were simply left as they are. Usually, the asymmetry and unevenness of these samples could not be significantly improved by further preparation. Solid machinable materials with different geometries were milled to the appropriate cross section and further ground to optimise the symmetry of the cross section and the surface planarity. Typically, the geometric quality of such samples was eventually around or below $10\upmu \mathrm{m}$.

Casting polymers such as epoxy, PDMS and polyurethane were cast bubble-free in a metal or PDMS mold. Solid samples were further ground to the precise shape, but elastomers were left with asymmetry and unevenness up to $50 \upmu \mathrm{m}$.

Materials available in foils and films were cut using a UV laser and laminated bubble-free using Araldite 2020 epoxy glue or Loctite 406 cyanoacrylate glue using rings of shrink tube as fixation. Both glues have a very low viscosity. We used a high-precision embossing press for thicker sheet material that machinable as a laminate whenever possible.
For glass and PCB samples, the thickness of the glue layers was not visible when observing a cut surface through the stack under the microscope and partially detached glass sheets showed widely-spaced interference rings. This suggests a $\upmu \mathrm{m}$- or sub-$\upmu \mathrm{m}$-scale thickness of the glue layer. A side view of stacked films also suggested at most a few $\upmu \mathrm{m}$ glue thickness and a comparison of the total thickness with the film thickness showed no glue thickness above uncertainties. As neither Araldite nor Loctite\footnote{The measurement of Loctite 406 failed twice due to air bubbles and problems with curing, but the susceptibility can be restricted to at most around $\pm 1$ ppm.} have an unusually large value of the susceptibility, we conclude that the glue layers are negligible.
%
%
%

Glass samples were cut using a wafer saw. Glass available in wafers was either first laminated and then cut, or first cut and then laminated. The former process was theoretically more accurate, but was problematic due to the limited adhesion strength on polished surfaces and the low glass transition temperature of the glue.
Up to 2.9 mm thick blocks were cut with the wafer saw in several steps. Here, the main problem was the repetition accuracy and softness of the blade, due to which there occurred trapezoidal errors to varying degrees in most samples -- leading to an accuracy around or below $10\upmu \mathrm{m}$.

Fluids were simply dispensed bubble-free into glass tubes that were sealed with epoxy on one side. On the other side, they were then first closed with cyanoacrylate glue and paper tissue (to accelerate curing), and finally sealed with epoxy. 
\subsection{Measurements}\label{measurements}
The phantoms for most samples were water-filled PET cylinders with 90 mm length and 62 mm diameter that were closed at the ends with PMMA.

Before proceeding with the final measurements, we performed a preliminary measurement to determine the parameters and the suitability of the methods, with samples of approximately $2\times 2\times 20 \, \mathrm{mm}^3$ at a spacing of 20 mm. Fluids were placed in glass tubes with 1.38 mm outer and 0.92 mm inner diameter.
These samples were mounted with acrylate glue on polyimide film in the PET cylinders, providing space for 13 samples per phantom. The water for this particular measurement contained an unknown concentration of contrast agent. As we used only two phantoms and the data had a very good quality, we could however compare the susceptibility values to the final measurement and estimate the shift due to the contrast agent to be 0.07 and 0.1 ppm, respectively. We found that the noise was very low such that there was no need for the contrast agent. The limiting factor was rather the mechanical accuracy of the samples.  After completing the final measurement, we also verified the final results against the preliminary ones\footnote{Fitting the overall shift took away only two degrees of freedom, out of overall 13 values.} and found a good agreement that reflected the statistics of the uncertainty distribution very well. 

\begin{figure}\begin{center}
\includegraphics[width=0.295\textwidth]{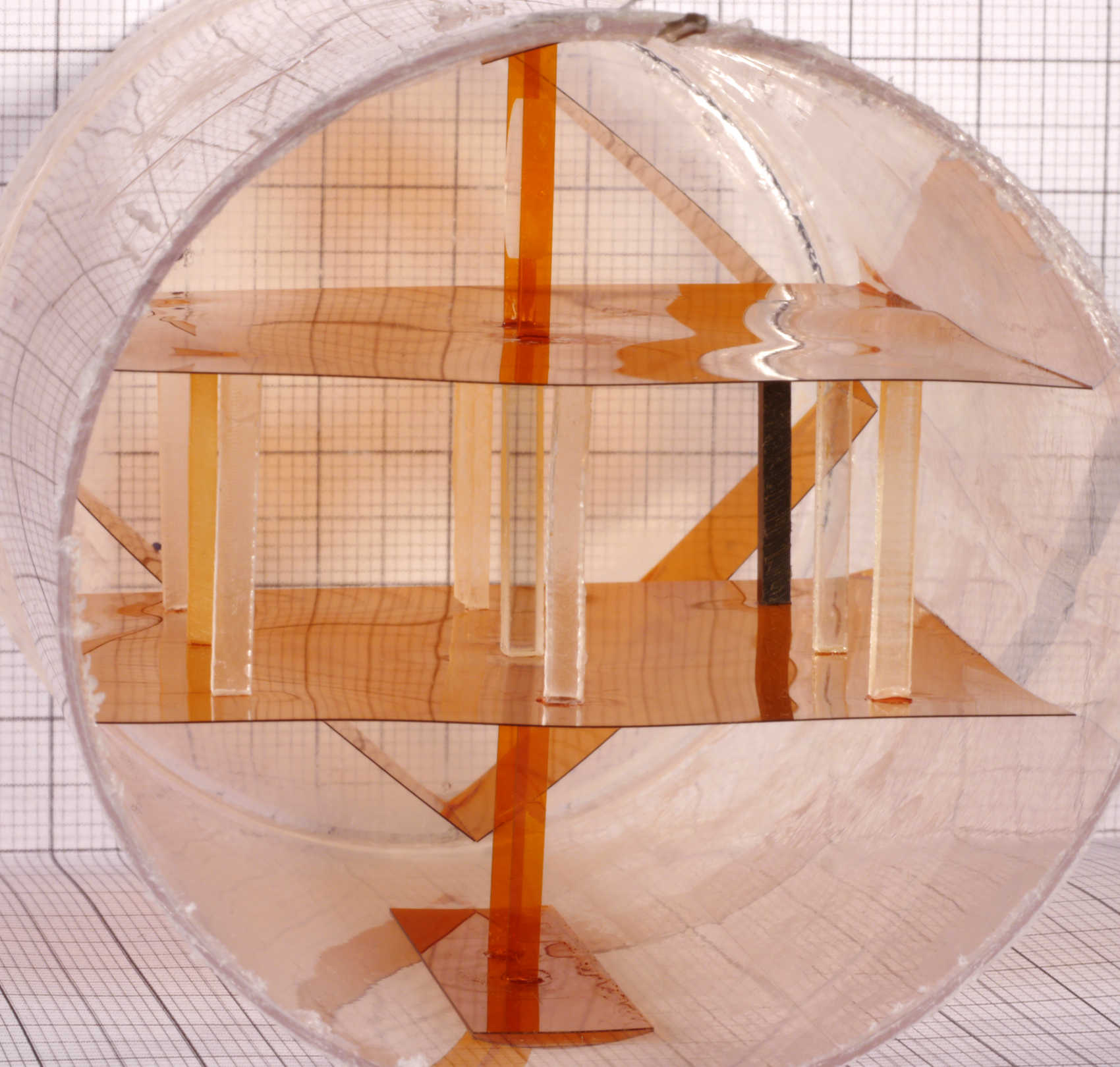}\hspace{0.01\textwidth}\includegraphics[width=0.301\textwidth]{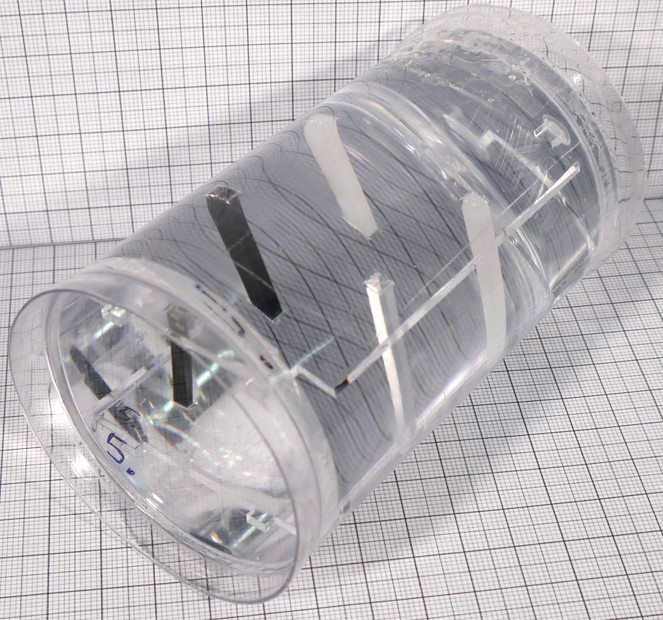}\hspace{0.01\textwidth}
\includegraphics[width=0.365\textwidth]{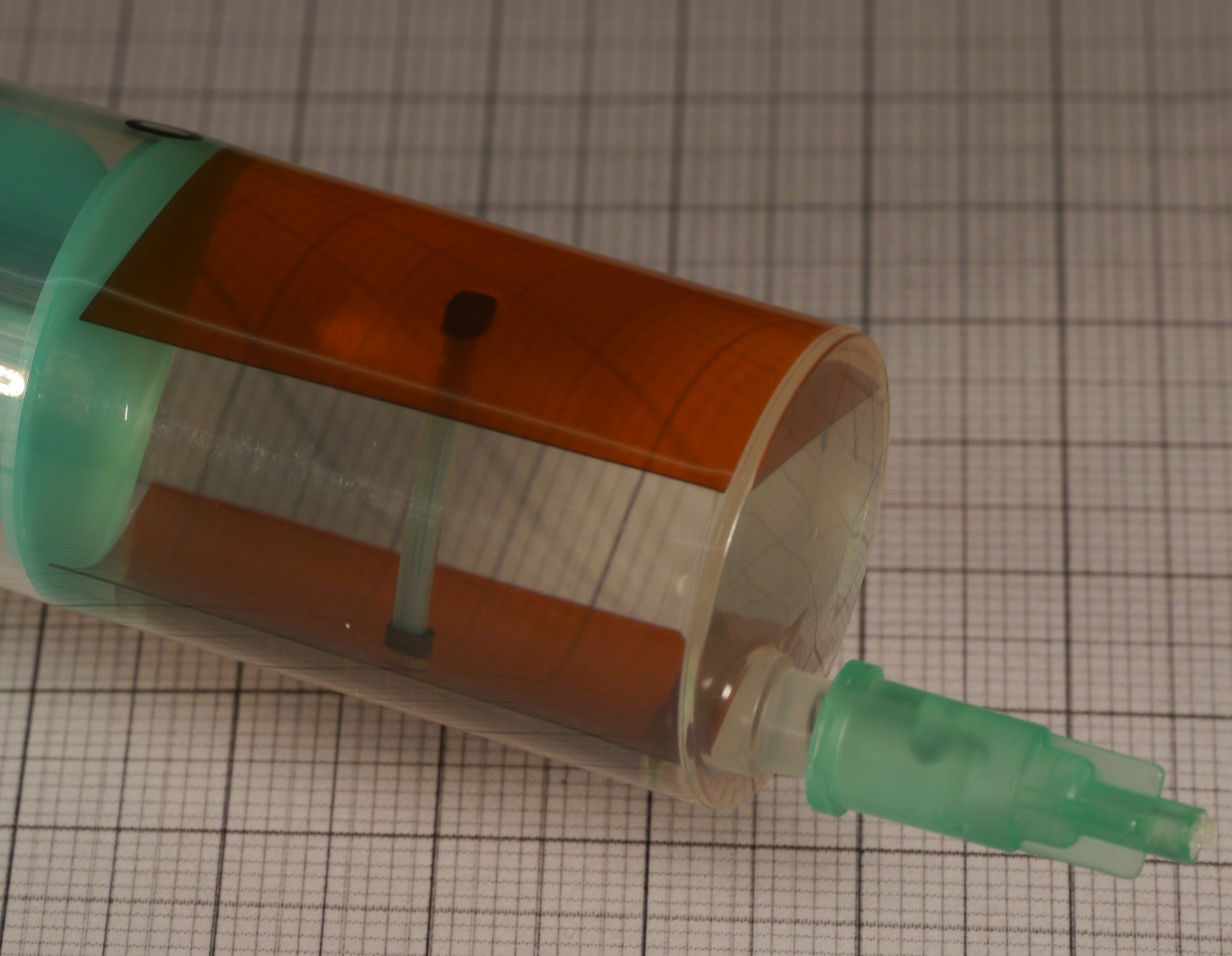}
\end{center}\caption{Photos of the phantoms, left to right: Preliminary measurement (cut open, without water), final measurement for large samples, final measurement for small samples.}\label{cylphot} 
\end{figure}
In the final measurements, samples available in bulk or in relatively large sizes were prepared with 30 mm length and a diameter of approximately 3 mm. These were mounted in an array of $2\times 2$ between two 2 mm thick PMMA sheets in the phantom cylinders, as shown in fig. \ref{cylphot}. The susceptibility of PMMA was shown in the exploratory measurement to be very close to water. Liquid samples were contained in glass tubes with an inner diameter of $2008 \pm 7\, \upmu \mathrm{m}$ and an outer diameter of $2780 \pm 7 \, \upmu \mathrm{m}$. These phantoms were filled with de-ionised water as a reference liquid, with a resistivity of $15 \, \mathrm{M}\Upomega\, \mathrm{cm}^{-1}$ at the time of de-ionisation, limited by dissolved $\mathrm{CO}_2$ and by the self-dissociation of water. 
%
Field maps were acquired with a 3D double gradient-echo sequence using the
following parameters: Repetition time (TR) 30 ms, echo times TE1 = 2.85 ms
and TE2 = 7.13 ms, flip angle = 15\degrees, field of view readout $\times$ phase
encoding1 $\times$ phase encoding2 =   $62 \times 62 \times 32.48 \mathrm{mm}^3$, matrix acquisition =
$210 \times 210 \times 110$. This allowed us to obtain an isotropic resolution of
$0.295 \times 0.295 \times 0.295 \, \mathrm{mm}^3$.

Small samples were mounted in 20 ml syringes made of polypropylene with an inner diameter of 20 mm and 32 mm length of the water volume, corresponding to 10 ml. Depending on their length, the samples were either mounted across the whole diameter of the syringe, or between two polyimide films. The reference liquid was also in this case de-ionised water. This measurement had a field of view of $21 \times 21 \times 21 \, \mathrm{mm}^3$ and a matrix acquisition $110 \times 110 \times 110$, giving a resolution of $0.191 \times 0.191 \times 0.191 \, \mathrm{mm}^3$. The sequence had TR = 40 ms, TE1 = 2.7 ms, TE2 = 6.98 ms and a flip angle of also 15\degrees. 
\subsection{Density measurements and absolute susceptibility}
As an approximately temperature-invariant parameter \cite{schenck1996role,haynes2012crc,philo1980temperature}, we would like to quote the total susceptibility relative to the density. Hence, we need the absolute value of the susceptibility of water and its temperature dependence. These, however, have been a long-standing issue of research \cite{schenck1996role,philo1980temperature,day1980equation,pan1970calculation,cini1968temperature,verhoeven1970magnetic,taft1969magnetic,das1959magnetic,arrighini1968magnetic}. To obtain the absolute value of the susceptibility, we use the value -9.032 ppm for the susceptibility of water at $20 \centigrade$ quoted in \cite{schenck1996role}, but we have to caution that our density-specific value critically hinges on this number.

Both to quote the density-specific susceptibility and also to identify polymers with varying composition and density, we measured the density for most of our samples, referenced to $20 \centigrade$. As most samples do not have perfectly regular surfaces we used immersion in de-ionised water to obtain the sample volume. The mass of the ``empty'' water-filled container gave us both the reference volume and an error estimate for the reproducibility and for the error distribution due to air bubbles and water droplets. For small samples of a few grams in a 4.5 ml container, we assumed 1 $\mathrm{mm}^3$ error, and for larger samples in a 55 ml container, we assumed 3 $\mathrm{mm}^3$ error, both of which agreed very well with the statistics of the ``empty'' water-filled weight. Further error sources that we took into account are the temperature variation relative to the temperature measurement during handling that we assumed to be $0.5 \centigrade$ and the variations in the thermal expansion coefficients that are only very roughly known from various usually unreferenced sources.
For the float glasses, we measured the density purely mechanically.

Where we used literature values for the density without accuracy figures, we assumed $\pm 1$ variation in the last digit when calculating volume-specific susceptibility but we do not state a variation for the density values itself in the tables. Certainly, this assumption may not always be correct. 
\section{Results}\label{results}
\subsection{Polymers, composites and PCB substrates}\label{polymers}
First, let us look at the results for polymers in table \ref{plastab}. 
\begin{table}\begin{center}
\begin{tabular}{|r||c|c|c|c|c|}
\hline
Material &  $\left( \chi - \chi_\mathrm{H_2 O}\right)/\mathrm{ppm}$ &  $\rho/(\mathrm{g/cm}^3)$ & $\frac{\chi/\mathrm{ppm}}{\rho/(\mathrm{g}\,\mathrm{cm}^{-3})}$ \\ \hline\hline
Fibreglass, R\&G $\ ^a$ & 5.92(2) & 2.082(3)& -1.49(1) \\ \hline
FR4 PCB board, Bungard $\ ^{a,b}$ & 5.29(2) & 1.952(2) & -1.918(9)\\ \hline
PZT Piezoceramics $\ ^c$ & 3.2 & -- & -- \\ \hline
FR3 PCB board $\ ^b$ & 1.361(7) & 1.601(2) & -4.792(6) \\ \hline
PET, Mylar A$\ ^d$ & 0.78(2) & 1.421(1) & -5.80(1) \\ \hline
Polyimide, Kapton $\ ^e$ & 0.117(8) & 1.42 \cite{kaptondata} & -5.57(4) \\ \hline
PMMA, clear & -0.023(4) & 1.1900(5) & -7.609(5) \\ \hline
PMMA, white & -0.036(8) & 1.195(1) & -7.59(1) \\ \hline
epoxy-based CRFP, R\&G DPP $\ ^a$ & -0.099(6) & 1.539(2) & -5.934(10) \\ \hline
Polystyrene & -0.141(9) & 1.045(1) & -8.78(2) \\ \hline
Polycarbonate & -0.210(5) & 1.1973(6) & -7.719(6) \\ \hline
Polycarbonate, Macrolon & -0.216(5) & 1.199(2) & -7.71(1) \\ \hline
PEEK & -0.300(6) & 1.3095(7) & -7.126(6) \\ \hline
POM, Sustarin C & -0.325(10) & 1.412(1) & -6.627(9) \\ \hline
POM & -0.342(5) & 1.417(2) & -6.617(8) \\ \hline
vinyl ester based CRFP, R\&G VEC  $\ ^a$ & -0.390(4) & 1.490(2) & -6.33(1) \\ \hline
PA6, natural $\ ^f$ & -0.446(5) & 1.1449(7) & -8.281(7) \\ \hline
Zeonex E48R & -0.505(6) & 1.008(1) & -9.46(1) \\ \hline
Polypropylene & -0.523(7) & 0.9110(9) &  -10.49(1) \\ \hline
PA, black & -0.532(4) & 1.1540(6) & -8.288(6) \\ \hline
PE-HD & -0.635(7) & 0.952(1) & -10.16(1) \\ \hline
FR2 PCB board $\ ^b$ & -1.080(5) & 1.352(3) & -7.48(1) \\ \hline
PTFE & -1.244(9) & 2.176(1) & -4.722(5) \\ \hline
PVC, gray $\ ^g$ & -1.673(8) & 1.3717(8) & -7.804(8) \\ \hline
\end{tabular}\\
{\footnotesize{ $\ ^a$ Fibres orthogonal to $B_0$; $\ ^b$ Without copper layer; $\ ^c$ Laminated from sheets with 110$\upmu m$ PZT with 64.1\% O, 8.6\% Ti, 2.7\% Sr, 9.6\% Zr and 15.0\% Pb atomar composition, unknown porosity and 5$\upmu m$ Silver electrodes of unknown composition, measured in the preliminary measurement; $\ ^d$ Laminated from $350 \upmu \mathrm{m}$ foils, parallel to $B_0$; $\ ^e$ Laminated from $120 \upmu \mathrm{m}$ foils, parallel to $B_0$; $\ ^f$ PA6.6 had a similar value of -0.4 ppm in the preliminary measurement; $\ ^g$ black PVC with a slightly higher density of 1.44 $\mathrm{g}\,\mathrm{cm}^{-3}$ had -1.8 ppm }}
\end{center}\caption{Susceptibility values for several polymers and polymer-based materials. The measurement uncertainties in the last digit (or sometimes last two digits) are shown in parentheses. Note the comments in the footnotes and in the text.}\label{plastab}
\end{table}
Looking at the extreme ends of very high and low susceptibilities, we have to caution that the realistic uncertainties of fibreglass, the PCB and possibly also PVC are probably larger than the ones shown here: There will be to some degree systematic errors coming from the geometric distortion of the field map due to the large magnetic field distortion, i.e. MR artifacts due to poor susceptibility matching. The susceptibility values of those former two epoxy-based glass fibre materials are somewhat surprising as neither glass nor epoxy are usually in the region of 5 ppm. If porosity were to account for that value, one would need more than 50\% air in the material, which is unlikely. A possibility is a high iron content in the glass used for the fibres. A similar high value appeared also in \cite{doty} for fibre-reinforced polyimide (A-JGN3030) with 2.5 ppm compared to water. Certainly one may think that fibre-based materials will have a large variation because of different possible compositions, but we can see that both of the each two carbon fibre and glass fibre materials are close together. Looking also at the other polymer materials from different manufacturers or even in differently coloured versions, we find that also they have all similar susceptibility values, significantly within 0.1 ppm.

One general observation is also that halogenated materials -- in this case PVC and PTFE -- tend to have a low susceptibility. This was also observed for PCTFE (Kel-F) with -2.6 ppm in \cite{doty} and for a perfluorinated oil in the next section. 

The value of the PZT material is only a rough indication as it was measured only in the preliminary measurement, the porosity of the material is unknown and the composition and processing of different PZT materials varies greatly. The value of 3 ppm cannot be accounted for only by air in the possibly porous material: By visual inspection of a polished sample, the porosity could be at most a few percent.
\begin{table}\begin{center}
\begin{tabular}{|r||c|c|c|c|c|}
\hline
Material &  $\left( \chi - \chi_{\mathrm{H_2 O}}\right)/\mathrm{ppm}$ &  $\rho/(\mathrm{g/cm}^3)$ & $\frac{\chi/\mathrm{ppm}}{\rho/(\mathrm{g}\,\mathrm{cm}^{-3})}$ \\ \hline\hline
RTV 615\footnotesize{ $\ ^a$ } & 1.00(2) & 1.0255(7) & -7.84(2) \\ \hline
Elastosil M 4642\footnotesize{ $\ ^a$ } & 0.93(1) & 1.1378(8) & -7.12(1) \\ \hline
Sylgard 184\footnotesize{ $\ ^a$ } & 0.927(9) & 1.031(1) & -7.86(1) \\ \hline
Makerbot PLA yellow\footnotesize{ $\ ^b$ } & 0.542(4) & 1.257(8) & -6.76(4) \\ \hline
Makerbot PLA red\footnotesize{ $\ ^b$ } & 0.538(4) & 1.256(2) & -6.76(1) \\ \hline
Makerbot PLA clear\footnotesize{ $\ ^b$ } & 0.535(5) & 1.252(2) & -6.79(1) \\ \hline
Makerbot PLA translucent blue\footnotesize{ $\ ^b$ } & 0.534(4) & 1.253(2)  & -6.78(1) \\ \hline
Makerbot PLA translucent green\footnotesize{ $\ ^b$ } & 0.530(4) & 1.255(2) & -6.78(1) \\ \hline
Makerbot PLA white\footnotesize{ $\ ^b$ } & 0.528(4) & 1.263(2) & -6.73(1) \\ \hline
Ordyl\footnotesize{ $\ ^c$ } & -0.265(5) & -- & -- \\ \hline
SUEX\footnotesize{ $\ ^c$ } & -0.59(1) & -- & -- \\ \hline
Epotec 201\footnotesize{ $\ ^a$ } & -0.611(7) & 1.160(1) & -8.31(1) \\ \hline
Araldite 2020\footnotesize{ $\ ^a$ } & -0.662(6) & 1.151(1) & -8.42(1) \\ \hline
SU8\footnotesize{ $\ ^d$ } & -0.7 & -- & -- \\ \hline
\end{tabular}\\
{\footnotesize{ $\ ^a$ Cast and cured from resin; $\ ^b$ 2 mm thick filament; $\ ^c$ laminated; $\ ^d$ obtained in the preliminary measurement }}
\end{center}\caption{Susceptibility values of silicones, PLA filament, photoresists and epoxies.}\label{gluetab}
\end{table}

Looking in table \ref{gluetab} at glues, photoresists, resins and 3D printing material we see again that the silicones at round 1 ppm and the epoxies around -0.6 ppm have only little variation in susceptibility. The PLA filament also shows that usually the effect of colouring additives is small.
\subsection{Liquids}\label{liquids}
The reference susceptibility value of the water in the two ``empty'' tubes vanishes
%
%
within errors at $0.00(1)\mathrm{ppm}$ and $-0.003(7)\mathrm{ppm}$. Similarly, the susceptibility values of the two reference tubes that were cut out of the same longer tube also agree within errors at $-2.12(1)\mathrm{ppm}$ and $-2.097(8)\mathrm{ppm}$.
\begin{table}\begin{center}
\begin{tabular}{|r||c|c|c|c|c|}
\hline
Material &  $\left( \chi - \chi_{\mathrm{H_2 O}}\right)/\mathrm{ppm}$ &  $\rho/(\mathrm{g/cm}^3)$ & $\frac{\chi/\mathrm{ppm}}{\rho/(\mathrm{g}\,\mathrm{cm}^{-3})}$ \\ \hline\hline
PEG 200 & 0.36(4) & 1.1240(3) & -7.72(4) \\ \hline
Paraffin oil 220-260 cts & 0.19(3) & 0.8780(3) & -10.07(4) \\ \hline
PBS buffer solution {\footnotesize{ $\ ^a$ }}& 0.009(9) & -- & -- \\ \hline
Ethylene glycol {\footnotesize{ $\ ^b$ }} &  0.0 & 1.1116(3) & -8.1 \\ \hline
Zeiss Immersol 2010 {\footnotesize{ $\ ^{b,c}$ }} &  -0.6 & 1.67 \cite{immersolw2010} & -5.8 \\ \hline
\end{tabular}\\
{\footnotesize{ $\ ^a$  Lonza Dulbecco's PBS with Calcium and Magnesium; $\ ^b$ measured in the preliminary measurement; $\ ^c$ A perfluoropolyether-alcohol}}
\end{center}\caption{Susceptibility values of liquids and gels.}\label{liqtab}
\end{table}

The results are shown in table \ref{liqtab}, where we list the values of polyethylene glycol, paraffin oil and phosphate buffer solution. A measurement of agarose gel failed because of microscopic air bubbles in an unusually high concentration of 10\% agarose, indicating however a value that might be slightly below the susceptibility of water. 
In the preliminary measurement, we further measured ethylene glycol and Zeiss Immersol 2010, a perfluoropolyether-alcohol used for optical immersion.
\subsection{Glass}\label{glasses}
The measurement results for various glass types are shown in table \ref{glasstab}.
\begin{table}\begin{center}
\begin{tabular}{|c|c|c|c|c|}
\hline
\multicolumn{5}{|l|}{Material, sample processing and issues} \\[-0.3em]
 $\left( \chi - \chi_{\mathrm{H_2 O}}\right)/\mathrm{ppm}$ &  $\rho/(\mathrm{g/cm}^3)$ & $\frac{\chi/\mathrm{ppm}}{\rho/(\mathrm{g}\,\mathrm{cm}^{-3})}$ & $n_d$ & $\nu_d$\\ \hline\hline
\multicolumn{5}{|l|}{AGC Anti-Newton, laminated from a 0.55 mm thick wafer} \\[-0.3em]
 2.963(9) & 2.49 \cite{AGCmanu} & -2.44(1) & 1.52 \cite{AGCmanu} & -- \\ \hline
\multicolumn{5}{|l|}{Schott N-SF57, cut from a 1 mm thick optical window} \\[-0.3em]
 0.645(7) & 3.53 \cite{schottopt} & -2.376(7) & 1.84666 \cite{schottopt} & 23.78 \cite{schottopt} \\ \hline
\multicolumn{5}{|l|}{Schott N-SF57, cut from 2.8 mm thick block} \\[-0.3em]
0.553(5) & 3.53 \cite{schottopt} & -2.402(7) & 1.84666 \cite{schottopt} & 23.7 8\cite{schottopt} \\ \hline
\multicolumn{5}{|l|}{Schott N-SF11, 1.5 mm diameter, ground from a lens} \\[-0.3em]
0.343(9) & 3.22 \cite{schottopt} & -2.698(9) & 1.78472 \cite{schottopt} & 25.68 \cite{schottopt} \\ \hline
\multicolumn{5}{|l|}{Schott N-SF10, cut from 2.8 mm thick block} \\[-0.3em]
 -0.284(5)& 3.05 \cite{schottopt} & -3.05(1) & 1.72828 \cite{schottopt} & 28.53 \cite{schottopt} \\ \hline
\multicolumn{5}{|l|}{Schott N-SF5, cut from 2.8 mm thick block} \\[-0.3em]
 -1.168(6) & 2.86 \cite{schottopt}  & -3.57(1) & 1.67271 \cite{schottopt} & 32.25 \cite{schottopt} \\ \hline
\multicolumn{5}{|l|}{Schott Borofloat B33, laminated from 0.5 mm glass wafers} \\[-0.3em]
 -2.024(5) & 2.22(1) & -4.98(2) & 1.47140 \cite{schottb33} & 65.20 \cite{schottb33} \\ \hline
\multicolumn{5}{|l|}{Schott AF32 eco, laminated from 0.7 mm glass wafers} \\[-0.3em]
 -2.15(1) & 2.43(1) & -4.60(2) & 1.5100 \cite{schottwafer} & 62.96 \cite{schottwafer} \\ \hline
\multicolumn{5}{|l|}{Schott Lithosil (quartz), laminated from 0.7 mm glass wafers} \\[-0.3em]
 -2.24(1) & 2.20(1) & -5.12(2) & 1.45843 \cite{schottlithosil} & 67.8 3\cite{schottlithosil} \\ \hline
\multicolumn{5}{|l|}{Schott D263T, laminated from 0.5 mm glass wafers} \\[-0.3em]
-2.560(6) & 2.49(1) & -4.66(2) & -- & -- \\ \hline
\multicolumn{5}{|l|}{Schott D263T eco, laminated from 0.7 mm glass wafers} \\[-0.3em]
 -2.693(8) & 2.49(1) & -4.71(2) & 1.5231 \cite{schottwafer}& 54.49 \cite{schottwafer}\\ \hline
\multicolumn{5}{|l|}{Schott B270, cut from a 1.3 mm thick optical window} \\[-0.3em]
-2.988(8) & 2.55(1) & -4.71(2) & 1.5230 \cite{schottB270}& 58.5 \cite{schottB270}\\ \hline
\multicolumn{5}{|l|}{Schott N-BK7 batch TJA 41562, cut from a 2.8 mm thick block} \\[-0.3em]
 -3.45(2) & 2.51 \cite{schottopt} & -4.97(2) & 1.51680 \cite{schottopt} & 64.17 \cite{schottopt} \\ \hline
\multicolumn{5}{|l|}{Schott N-BK7 batch D 2044011, cut from a 2.8 mm thick block} \\[-0.3em]
-3.59(1) & 2.51 \cite{schottopt} & -5.03(2) & 1.51680 \cite{schottopt} & 64.17 \cite{schottopt} \\ \hline
\multicolumn{5}{|l|}{Schott N-LaK8, cut from a 2.8 mm thick block} \\[-0.3em]
 -4.97(1) & 3.75 \cite{schottopt} & -3.74(1) & 1.71300 \cite{schottopt} & 53.83 \cite{schottopt} \\ \hline
\multicolumn{5}{|l|}{Schott SF11, cut from a 1mm thick optical window} \\[-0.3em]
 -5.37(5) & 4.47 \cite{schottopt} & -3.22(1) & 1.78472 \cite{schottopt} & 25.76 \cite{schottopt} \\ \hline
\end{tabular}
\end{center}\caption{Susceptibility values of various glasses. From the results obtained for different batches for N-BK7 and N-SF57, one should assume a higher uncertainty of $\pm 0.1 \mathrm{ppm}$. For the large deviations from water in the region of 3 ppm and above, there comes an additional uncertainty from the distortion of the field map, that was not quantified.}\label{glasstab}
\end{table}
We find that there is a large variation of susceptibilities over different glass types. For the extremely low susceptibilities, this does not surprise because of the high densities, even though the absolute value of the susceptibility does not scale proportional to the density. At the higher susceptibilities, we see the other effect, that eventually the glass materials are just metal oxides with varying compositions. The very high susceptibility of the AGC Anti-Newton soda lime glass, for example, may be due to its iron content of $0.08\%$ $\mathrm{Fe_2 O_3}$ by weight \cite{AGCmanu}.

Furthermore there is also a variation between different batches, significantly greater than the measurement uncertainties. This is  surprising, as it cannot be attributed to measurement uncertainties, in particular since we saw much smaller variations for polymers. At least for the optical glasses the manufacturing process should be very consistent. For glasses containing small amounts of iron, this might be due to a high sensitivity of the susceptibility to the iron concentration. Because of these variations, one should assume an uncertainty around $\pm 0.1\, \mathrm{ppm}$ rather than the stated values. Furthermore, the ``old'' types of the optical glasses can be very different from the ``new'' lead-free versions, as we saw in SF11 versus N-SF11.
\begin{figure}\begin{center}
\includegraphics[width=0.31\textwidth]{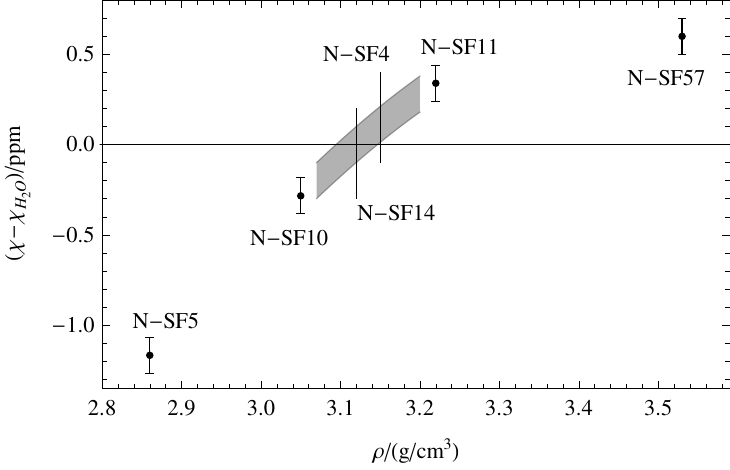}\hspace{0.01\textwidth}
\includegraphics[width=0.31\textwidth]{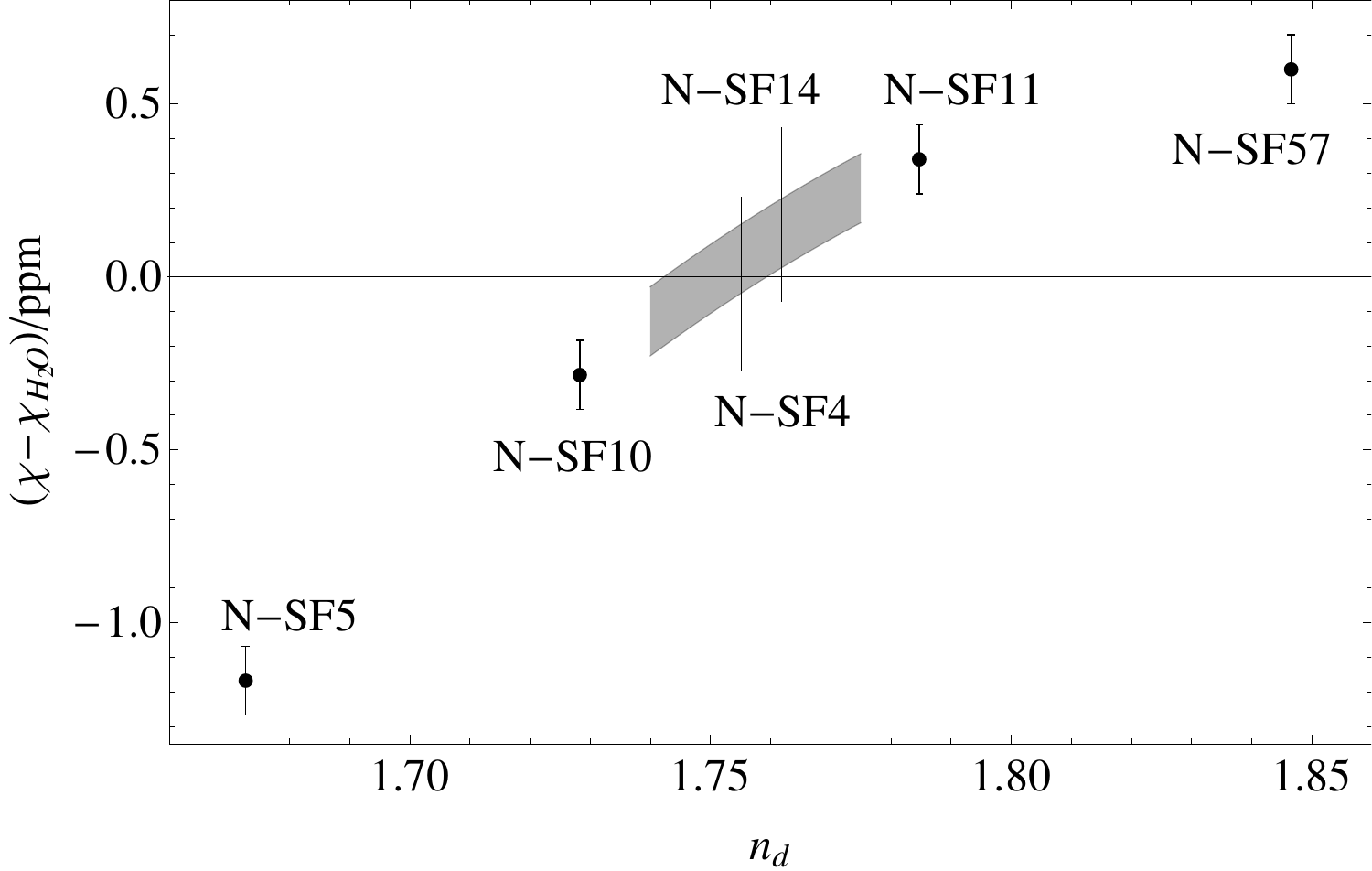}\hspace{0.01\textwidth}
\includegraphics[width=0.31\textwidth]{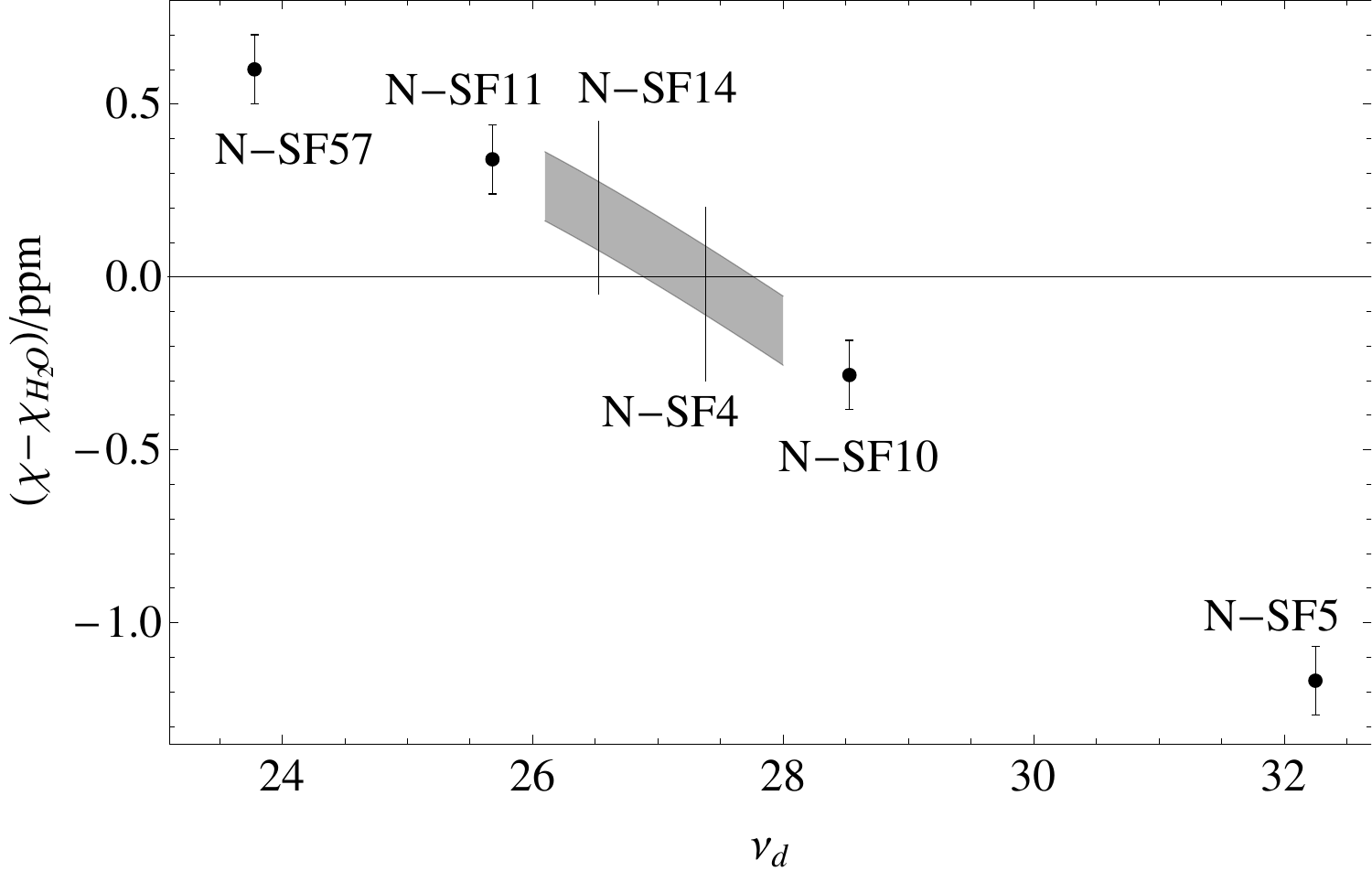}\hspace{0.01\textwidth}
\end{center}\caption{The susceptibility of dense flint glasses depending on the density, refractive index and Abbe number. The gray shaded area indicates which susceptibility other glasses might have according to the present trend. N-SF4 and N-SF14 are indicated with black lines.}\label{glassfig} 
\end{figure}

For most glasses, there is no clear dependence of the susceptibility on the other material properties. The dense flint glasses, however, seem to have a trend that is shown in fig. \ref{glassfig}. While these are optical glasses that are mostly used for lenses and prisms, it is still interesting to see how close they can be to water. N-SF10 and N-SF11  still have a difference to water around $0.3 \, \mathrm{ppm}$ but there are two candidates, N-SF4 and N-SF14, that could be a perfect substrate material for micro-MR.
\subsection{Polyurethane}\label{polyurethanes}
Polyurethanes are a huge class of materials. Here, we concentrate on various transparent 2-component resins.

Since PDMS (silicone) gives an MR signal close to the one of water, we investigated elastic PU as an alternative. Hence, we recorded also the spectra in addition to the susceptibility. As PU showed a susceptibility close to water in the preliminary measurements and it can be shaped either by casting and curing the resin or by mechanical processing, we also studied hard polyurethane. As part of this investigation, we considered different curing parameters, different mixtures of the resin and various functional additives.
\begin{table}\begin{center}
\begin{tabular}{|r||c|c|c|c|c|}
\hline
Material &  $\left( \chi - \chi_{\mathrm{H_2 O}}\right)/\mathrm{ppm}$ & $T_{\mathrm{curing}} $  & $\rho/{(\mathrm{g/cm}^3)}$ & $\frac{\chi/\mathrm{ppm}}{\rho/(\mathrm{g}\,\mathrm{cm}^{-3})}$ \\ \hline\hline
Polytek GlassRub 50; 1A:1B & 0.313(5) & 43$^\circ \mathrm{C}$  & 1.0165(7) & -8.577(8) \\ \hline
Polytek GlassRub 50; 10A:8B & 0.289(8) & 43$^\circ \mathrm{C}$  & 1.018(1) & -8.58(1) \\ \hline
Polytek GlassRub 50; 10A:12B & 0.275(8) & 43$^\circ \mathrm{C}$  & 1.0167(10) & -8.61(1) \\ \hline
BJB WC-540 & 0.094(5) & 38$^\circ \mathrm{C}$  & 1.068(2) & -8.37(2) \\ \hline
2 WC-540 : 3 WC-565 & 0.033(7) & 38$^\circ \mathrm{C}$  & 1.079(4) & -8.34(3) \\ \hline
3 WC-540 : 2 WC-565 & 0.025(8) & 38$^\circ \mathrm{C}$  & 1.071(3) & -8.41(2) \\ \hline
BJB WC-565 & 0.007(6) & 38$^\circ \mathrm{C}$  & 1.0797(8) & -8.359(8) \\ \hline
Smooth-On Clear Flex 50 & -0.026(6) & 28$^\circ \mathrm{C}$  & 1.089(2) & -8.32(2) \\ \hline
Smooth-On Clear Flex 50 & -0.028(5) & 38$^\circ \mathrm{C}$  & 1.0875(7) & -8.331(7) \\ \hline
Smooth-On Clear Flex 50 & -0.031(5) & 53$^\circ \mathrm{C}$  & 1.086(1) & -8.344(10) \\ \hline
Clear Flex 50 10A:18B & -0.061(10) & 38$^\circ \mathrm{C}$  & 1.092(2) & -8.32(2) \\ \hline
Clear Flex 50 10A:15B & -0.113(6) & 38$^\circ \mathrm{C}$  & 1.098(1) & -8.328(10) \\ \hline
Polytek Poly-Optic 1470 & -0.098(6) & 43$^\circ \mathrm{C}$  & 1.1085(7) & -8.237(8) \\ \hline
Poly-Optic 10$\times$1470:2$\times$1410 & -0.19(2) & 43$^\circ \mathrm{C}$  & -- & -- \\ \hline
Poly-Optic 10$\times$1470:5$\times$1410 & -0.215(6) & 43$^\circ \mathrm{C}$  & 1.120(2) & -8.26(2) \\ \hline
Polytek Poly-Optic 1410 & -0.249(5) & 43$^\circ \mathrm{C}$  & 1.1209(8) & -8.280(7) \\ \hline
P-O 1410 w. 2.0\% P-O 14R & -0.25(1) & 43$^\circ \mathrm{C}$  & 1.119(1) & -8.30(2) \\ \hline
1 P-O 1410 : 1$\times$ BJB WC-783 & -0.251(4) & 43$^\circ \mathrm{C}$  & 1.112(1) & -8.35(1) \\ \hline
BJB WC-783 & -0.255(5) & 43$^\circ \mathrm{C}$  & 1.1053(7) & -8.402(7) \\ \hline
1 CC 204 : 1 BJB WC-783 & -0.245(5) & 43$^\circ \mathrm{C}$  & 1.107(1) & -8.38(1) \\ \hline
1 CC 204 : 1 Poly-Optic 1410 & -0.240(6) & 43$^\circ \mathrm{C}$  & 1.114(1) & -8.32(1) \\ \hline
Smooth-On Crystal Clear 204 & -0.233(4) & 53$^\circ \mathrm{C}$  & 1.108(1) & -8.36(1) \\ \hline
Smooth-On Crystal Clear 204 & -0.246(8) & 33$^\circ \mathrm{C}$  & 1.109(2)& -8.36(1) \\ \hline
Crystal Clear 204, 36h curing & -0.233(4) & 43$^\circ \mathrm{C}$  & 1.108(1) & -8.36(1) \\ \hline
Crystal Clear 204, 12h curing & -0.244(8) & 43$^\circ \mathrm{C}$  & 1.1096(7) & -8.360(9) \\ \hline
Crystal Clear 204, 8h curing &  -0.241(4) & 43$^\circ \mathrm{C}$  & 1.114(1) & -8.320(10) \\ \hline
CC 204 with 0.9\% BYK-054 & -0.232(8) & 43$^\circ \mathrm{C}$  & 1.107(1) & -8.37(1)  \\ \hline
CC 204 with 1.8\% BYK-1152 & -0.224(6) & 43$^\circ \mathrm{C}$  & 1.106(1) &  -8.37(1) \\ \hline
CC 204 with 0.8\% BYK-141 & -0.249(8) & 43$^\circ \mathrm{C}$  & 1.109(1) & -8.37(1)  \\ \hline
Alchemix PU 3660 & -0.303(9) & --  & 1.118(1) & -8.35(1)  \\ \hline
\end{tabular}
\end{center}\caption{Susceptibility values of various soft and hard polyurethanes, in different mixtures, with additives and with different curing conditions. The temperature and curing time are accurate to $\pm 2^\circ \mathrm{C}$ and $\pm 2\, \mathrm{h}$, respectively. Where no curing time is given, the time was typically 8 to 16 hours. The Alchemix PU 3660 is a professionally cured sample from the manufacturer.} \label{putab}
\end{table}

In table \ref{putab}, we find first of all that all polyurethanes have a susceptibility close to the one of water. The curing temperature seems to have no significant influence on the susceptibility value, whereas longer curing seems to lower both the density and susceptibility -- the latter still below the threshold of significance. If we compare different polyurethanes of the same family, BJB Water Clear 565 and 540, Polytek Poly-Optic 1410 and 1470 or the different mixtures of Smooth-On Clear Flex 50, we see that softer polyurethanes seem to have higher susceptibility values. Still, the hardness is no general indicator for the susceptibility, as Clear Flex 50 with shore hardness A50 has a lower susceptibility than Water Clear 565 with shore hardness A65. The mixtures of different types of polyurethanes were produced by first mixing the components and then the uncured mixtures. We find that the susceptibility seems to reflect the mixing, but there is no linear relation to the mixing ratio. The different additives -- the curing retarder Poly-Optic 14R and the defoamers from BYK -- have little to no influence on the susceptibility. This is not surprising as the retarder is just a modified resin component and the additives are just some silicones and other polymers, with susceptibilities usually not very far away from water. A serious potential error source might be the diffusion of water into the material. We measured 4\% and 1.5\% long-term absorption of water by submerged soft Clear Flex 50 and hard Crystal Clear 204, respectively. As the susceptibility values are very close to water, however, we do not expect a shift in susceptibility significantly greater than our error estimates. We found similar absorption by epoxy, and also for example Polyimide is known to have such a level of water absorption. 

We will comment about the somewhat unusual Polytek GlassRub below, in the context of the spectra.

%
The spectra were measured in a 7 T Bruker Biospec 70/20 imaging system, also with maximum gradient amplitude of 676 mT/m and 4750 mT/m/s maximum slew rate.
To obtain a reliable signal of the spectra above the backgrounds of the MR scanner, we produced relatively large cylindrical samples with a diameter of 26 mm and a weight of 15-18 g that we measured in a quadrature volume coil with inner diameter of 40 mm. The spectra were obtained using a non-localised single acquisition pulse-and-acquire FID sequence with 90\degrees flip angle and a delay of t = 2 ms or 3 ms between excitation and begin of acquisition. Acquisition duration was 273 ms, 2048 sample points were recorded.
\begin{figure}\begin{center}
\includegraphics[width=\textwidth]{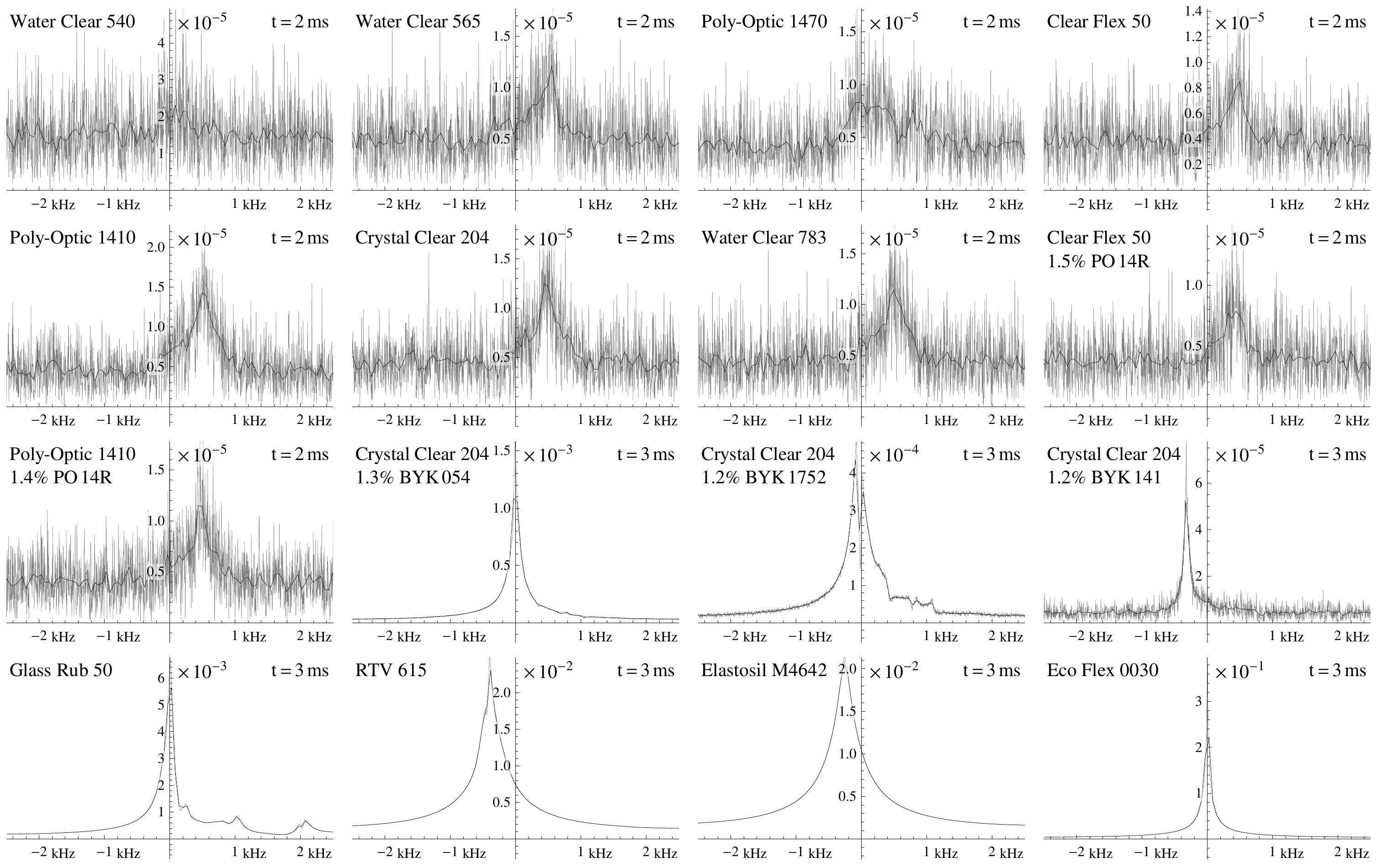}
\end{center}\caption{The magnitude spectra of a collection of soft (first row) and hard (first three of second row) polyurethanes, polyurethanes with different additives and PDMS (last three of last row) for comparison. The SNR is normalised to the one of water, weighted by the sample mass. The gray curve shows a resolution of 3.67 Hz and the black curve shows a re-sampling over 16 values, i.e. a resolution of 58.6 Hz.}\label{spectra} 
\end{figure}

The spectra of a collection of hard and soft polyurethanes, polyurethane with curing-retarder and de-foaming additives and, for comparison and three PDMS samples can be found in fig. \ref{spectra}. To present the data, we normalised the spectra by sample weight relative to the SNR of water, i.e. water has a peak value of 1. As expected, we see that PDMS produces a large signal. For polyurethane, there is a very small signal in most materials. The only exception were samples doped with the silicone-based de-foaming agents BYK-054 and BYK-1752, and the elastomer GlassRub 50. The latter also had an unusual susceptibility of $0.313(4)\mathrm{ppm}$ and an unusual brown-tinted appearance. Hence, one can assume that it is heavily modified. 
\begin{table}\begin{center}
\begin{tabular}{|r||c|c|}
\hline
Material & Shore hardness & Total signal \\ \hline\hline
BJB WC-540 & A40 & $16\times 10^{-6}$   \\ \hline
BJB WC-565 & A65 & $13\times 10^{-6}$   \\ \hline
Polytek Poly-Optic 1470 & A70 & $14\times 10^{-6}$   \\ \hline
Smooth-On Clear Flex 50 & A50 & $7\times 10^{-6}$   \\ \hline
BJB WC-783 & D82 & $15\times 10^{-6}$   \\ \hline
Polytek Poly-Optic 1410 & D85 & $21\times 10^{-6}$   \\ \hline
Smooth-On Crystal Clear 204 & D80 & $19\times 10^{-6}$   \\ \hline
Clear Flex 50, 1.5\% Poly-Optic 14R & A50 & $7\times 10^{-6}$   \\ \hline
Poly-Optic 1410, 1.5\% Poly-Optic 14R  & D85 & $14\times 10^{-6}$   \\ \hline
Crystal Clear 204, 1.3\% BYK-054 & D80 & $1.6\times \mathbf{10^{-3}}$   \\ \hline
Crystal Clear 204, 1.4\% BYK-1752 & D80 & $0.9\times \mathbf{10^{-3}}$   \\ \hline
Crystal Clear 204, 1.3\% BYK-141 & D80 & $40\times 10^{-6}$   \\ \hline
Polytek GlassRub 50 & A50 & $8\times \mathbf{10^{-3}}$   \\ \hline
RTV 615 & A44 & $0.054$   \\ \hline
Elastosil M 4642 & A37 & $0.057$   \\ \hline
Eco Flex 0030 & OO30 & $0.25$   \\ \hline
\end{tabular}
\end{center}\caption{The integrated spectrum of various soft and hard polyurethanes, polyurethanes with different additives and PDMS for comparison. The spectra are normalised by weight to the integrated spectrum of water, and the shore hardness is taken from manufacturer documentation. The A-scale is used for soft material like rubber, D-scale for hard materials like plastics and OO-scale for very soft materials like gels.}\label{spectab}
\end{table}

To numerically compare the materials, we show the integrated spectrum in table \ref{spectab}, normalised by weight to the integrated spectrum of water. 
We find that the signal of polyurethane is very small. Polyurethane without de-foaming additives has a signal by a factor of $10^{-5}$ smaller than water, and around $10^{-3}$ smaller than PDMS -- actually independent of the softness of the material. This signal may come either from the protons in the material itself or from water absorbed into the material or other contaminations. 

As one might in fact worry here about signal coming from absorbed water, we measured again the absorption of water, this time by soft ClearFlex 50 from the air humidity. This gave 0.6\% to 0.8\% weight gain in typical laboratory conditions of 35\% to 65\% relative humidity and 22\degrees C to 25 \degrees C compared to a nitrogen atmosphere. This means that even if all signal of the sample comes from water, the signal of the water in the material is suppressed by a factor of 50 compared to liquid water. This does not come as a surprise, as bound water is known to have a short relaxation time and we measured the spectra at 2 ms.
%
%
\section{Comparison to the literature}\label{comparison}
Before we summarise, we compare our results to known values of identical or equivalent materials from  the most comprehensive measurements of engineering materials in the literature, \cite{keyser1989magnetic} and \cite{doty}. As their results are less accurate than ours, we further verified the consistency and reproducibility by comparison to our exploratory measurement as described in section \ref{measurements}.
\begin{table}\begin{center}
\begin{tabular}{|r||c|c|c|c|}
\hline
Material &  $\chi_{\mathrm{meas.}} /\mathrm{ppm}$ &   $\chi_\mathrm{lit.} /\mathrm{ppm}$ \cite{keyser1989magnetic} &   $\left( \chi_{\mathrm{meas.}} -\chi_\mathrm{lit.} \right) /\mathrm{ppm}$ & $\chi_\mathrm{lit.} /\mathrm{ppm}$ \cite{doty} \\ \hline\hline
RTV 615  & -8.03(2) &  -8.32(31) & 0.31(0.31) & -- \\ \hline
``Silicone'' & -- & -- & -- & -7.8 \\ \hline
PI, Kapton & -8.917(8) & -- & -- & --  \\ \hline
PI, Vespel SP1/Vespel & -- &  -9.02(25) & 0.10(25) & -9.2 \\ \hline
Ethylene glycol & -9.0 &  -- & -- & ``$\sim \, -5$'' \\ \hline
PC, Macrolon  & -9.248(5) &  -- & -- & -- \\ \hline
PC, Lexan  & -- &  -9.56(25) & 0.31(25) & -- \\ \hline
PEEK & -9.332(6) & -- & -- & 9.3 \\ \hline
POM, white & -9.372(5) $\ ^a$ &  -- & -- & -- \\ \hline
POM, Delrin & -- &  -9.48(25) & 0.11(25) & -- \\ \hline
PTFE & -10.276(9) &  -10.25(25) & -0.03(25) & -10.5 \\ \hline
SCHOTT Borofloat B33 &  -11.056(5) & -- & -- &  --\\ \hline
Pyrex-7070 & -- & -- & -- & -11.0 \\ \hline
Quartz, Lithosil & -11.27(1) &  -- & -- & -- \\ \hline
``Fused quartz''/``Quartz'' & -- &  -11.94(25) & 0.67(25) & -11.8 \\ \hline
N-BK7 & -11.52(7) &  --  & -- & --  \\ \hline
``BK7'' & -- &  -10.63(25) & -1.89(26) & -- \\ \hline
\end{tabular}\\
{\footnotesize{ $\ ^a$ weighted mean}}
\end{center}\caption{Comparison of our measurements (left column) to the results of \cite{keyser1989magnetic} and their difference (central two columns) and to the results of \cite{doty}. The reader should keep in mind the uncertainty of typically $\pm 0.25 \mathrm{ppm}$ in the central columns, the uncertainty in the right column is unknown.}\label{compltab}
\end{table}

The method of \cite{keyser1989magnetic} was based on a force measurement of the sample in a magnetic field, so there may be no common methodological problems affecting both measurements in the same way.
To compare the values, we added the literature value of the susceptibility of water at 20\degrees C, -9.032 ppm \cite{schenck1996role}, to our results, assuming that the uncertainty of that value is negligible compared to our accuracy. The the comparison to \cite{keyser1989magnetic} in table \ref{compltab} show a good agreement within the uncertainties of their measurement, except for the value of N-BK7 vs. BK7. We presume that what is referred to as BK7 in \cite{keyser1989magnetic} is indeed the ``old style'' BK7 or another glass with the same optical properties from a different manufacturer, not N-BK7, and may thus have a significantly different susceptibility as we also observed in the dense flint glasses. 

As described in section \ref{methods}, the measurement of \cite{doty} measures the frequency shift in an NMR sample to obtain the field distortion from the sample material placed in a Helmholtz configuration around the NMR sample at 7 T. Hence this method is more similar to ours. As they do not state exact error figures, we cannot do a quantitative statistical comparison. Assumed that the accuracy is of the order of 0.1 ppm or a few 0.1 ppm as suggested by the last stated digit, also this comparison shows good agreement -- with a deviation of at most 0.5 ppm. The value for ethylene glycol in \cite{doty} is probably not correct. In \cite{kuchel}, the authors calculated a value of -8.77 ppm (without an error figure) from the molar susceptibility value in \cite{haynes2012crc}, which is reasonably close to our value, given that the uncertainty in our preliminary measurement was around 0.05 ppm. This is also close to our value for PEG 200 in section \ref{liquids} and to other organic fluids of a similar density in \cite{doty}.
\section{Summary, discussion and conclusions}\label{conclusions}
\subsection{Methods and accuracy}
We implemented a reliable method for the measurement of the magnetic susceptibilities of diamagnetic materials relative to the susceptibility to water. This was used to produce an overview of the susceptibility values of several materials that may be of particular interest for MR applications. One focus was placed on polyurethanes, where we also measured the MR spectra in regard to MR compatibility; another focus was placed on glass materials.

The susceptibility was obtained from the distortion of a homogeneous magnetic field caused by a material sample in a water-filled phantom as described in section \ref{methods}. This homogeneous magnetic field was the 9.4 T $B_0$ field of a Bruker BioSpec 90/21 small animal scanner and we obtained the field distortion by measuring the field map.
Solid samples were prepared as rods with quadratic or circular cross sections and fluids were measured inside small glass tubes, all oriented orthogonal to the $B_0$ field and with similar aspect ratios. The standard sample size was $3\times 3\times 30 \, \mathrm{mm}^3$ but also smaller samples down to 1 mm diameter were used in some cases where the materials were not available in the appropriate size.
The susceptibility value was then obtained by performing a linear fit of the measured field distortion slice-by-slice to the field distortion in a 3D finite element simulation. 
As sources for the presented error figure, we included the geometric accuracy of the samples, the accuracy of the fit, the variation of the obtained susceptibility values over the length of the bar, the thermal variation and the variation between different models for fitting the ``background'' field inhomogeneity in the phantom and different sizes of the volume of interest. Additional possible error sources that were not included in the uncertainty figure were commented about in the results in section \ref{results}. 

To our knowledge, none of the tested materials shows a ferromagnetic or antiferromagnetic behaviour, and all the tested materials were diamagnetic with a susceptibility of the order of $-10^{-5}$. Hence, we believe that the results also apply to other magnetic fields. This is supported in the case of some materials in section \ref{comparison} from literature data, which was taken at other flux densities. 

For susceptibilities close to water, the uncertainties of the measurement results in section \ref{results} were 
often only limited due to thermal expansion of water to $\pm 0.004 \, \mathrm{ppm}$. As the temperature variation in most MR measurements, in particular in small micro-MR probes, will not be much smaller than our $\pm 2 \, ^\circ \mathrm{C}$, this accuracy should not pose a constraint when using the data. Our accuracy is by one to two orders of magnitude better than the largest previous literature data sets for engineering materials \cite{doty,keyser1989magnetic}.

For values of several ppm away from water, we expect, however, that the accuracy drops due to geometric distortions of the measured field map. The variation of different materials of the same type was for glasses in the range of 0.1 ppm and for polymers usually significantly below -- around a few 0.01 ppm. This indicates that our results may be to some level universal, also for materials of the same type from different manufacturers or slightly varying composition. While we measured only few liquids presented in section \ref{liquids}, the method is also accurate for liquids. In particular we can also measure liquids that are not based on water. This method is also suitable to measure powders and small particles or filaments in suspension in a suitable solvent by comparing a measurement of the solvent to a measurement of the suspension with known concentration.

We measured only non-conductive materials\footnote{In the carbon fibre materials, the eddy currents should be negligible due to the fibre orientation and the relatively low conductivity of carbon.}. As obtaining the field map requires a gradient field, conductive sample materials would have an induced eddy current. The resulting heating and vibrations need to be taken into consideration for such materials and may limit this measurement method.
\subsection{Results}
The results for polymers in section \ref{polymers} indicate that there are several materials with various physical properties available below $\pm 0.3$ ppm susceptibility difference to water.
The probably most MR-compatible material is PMMA. For applications where autoclaving is needed, polyimide may be suitable, but one has to keep in mind possible water absorption. Alternatively PEEK and POM may be a choice; or polystyrene and polycarbonate in cases where manufacturing aspects, cost or optical transparency are more important than mechanical and thermal properties. PTFE, while it intrinsically does not give an $H^1$ signal, may cause magnetic artifacts due to its susceptibility, as may also PVC and perfluoropolyether oil. Particularly unsuitable materials were glass fibre laminates. As long as the conductivity of carbon causes no problem, they might be replaced by carbon fibre, or for PCBs by FR2 rather than FR4. The one PZT material had a susceptibility 3.2 ppm higher than water, which is still safe provided it is not arranged close to the sample.

Another polymer with a good susceptibility was polyurethane, which has a wide range of applications e.g. as prototyping or casting resin, glue or elastomer. In the survey of transparent PU in section \ref{polyurethanes}, we found that the hard polyurethanes have  $\left( \chi -\chi_\mathrm{H_2 O} \right)$ between -0.3 ppm and -0.2 ppm, usually near -0.25 ppm, and most soft ones have $\left| \chi -\chi_\mathrm{H_2 O} \right| < 0.1 \, \mathrm{ppm}$. As PDMS, a commonly used elastomer, does give an MR signal close to water, we measured also the spectra of PU and found that with one exception, the signal of PU is smaller than water by a factor of $10^{-5}$ at an echo time of 2 ms. At longer echo times, no signal above noise was visible. We also looked at different processing conditions, which have little influence on the susceptibility.

In our survey of different glass types of section \ref{glasses}, we found a large variation in the susceptibility values. However, we found two types of optical glass, Schott N-SF11 and N-SF10 that are only 0.3 ppm away from water, making them in principle suitable for applications in micro-MR. There seems to exist a trend in the susceptibility of dense flint glasses depending on other physical parameters, such that one can identify further candidates that might be even closer to water. This trend is not surprising as these materials are probably obtained by incrementally varying their composition and processing parameters.

Over all classes of materials, we found one common feature, that the total susceptibility varies much less than the density. 

Finally, we compared our results to literature values for those materials where that was possible and found good agreement. We further compared the final measurement to results from an exploratory experiment which verifies the reproducibility and the consistency of our results and uncertainty estimates.
\section*{Acknowledgements}
We would like to thank Nicoleta Baxan and Michael Herbst for help with the MR measurements, Frederik Testud for useful discussions and hints for the literature and Moritz St\"urmer for the EDX measurement of the PZT ceramics.

We are grateful for the companies SCHOTT AG Advanced Optics, Zeon Chemicals and Alchemie Ltd. for providing us free samples of their materials and Gudrun St\"ander from SCHOTT for arranging the samples and supplying data sheets.

This research was financed in part by the Baden W\"urttemberg Stiftung through the project ADOPT-TOMO and supported partly by BrainLinks-BrainTools Cluster of Excellence funded by the German Research Foundation (DFG, grant number EXC 1086).

\bibliography{mribib}
\bibliographystyle{model1-num-names}

\end{document}